# Waveform discrimination by fitting derivative of synchronized ideal normalized curves, dSINC fit

Martin Andersson[a*], Tatsuo Torii[b], Susumu Ryufuku[a], Ryohei Kurosawa[a], Hiroko Kido[a]

[a] Visible Information Center, Inc., 440, Muramatsu, Tokai-mura, Naka-gun, Ibaraki, 319-1112, Japan;
[b] Institute of Environmental Radioactivity, Fukushima University 1, Kanayagawa, Fukushima-shi, Fukushima, 960-1296, Japan

**ABSTRACT**

dSINC proposes an alternative algorithm for waveform discrimination of measurement data from multi-layer scintillator sandwich designs. dSINC attempts to solve problems related to noise and peaks-piling sensitivity in the feature extraction step of traditional KNN waveform discrimination, by fitting the derivative of the entire gain section of the waveform against ideal waveforms learned from training data – and thereby completely sidestepping the problems of feature extraction.

[*]Corresponding author. Email: martin@vic.co.jp

# 1 Introduction

A detector that combines two or more scintillators with different optical properties and a photomultiplier tube or other optical-to-electrical signal conversion device is called a phoswich (or phosphor sandwich) detector. This detector is used to discriminate the type of incident radiation. In particular, scintillators with different decay times are used and combined so that the shape of the signal pulse output from the conversion device depends on the relative contribution of scintillation light from two or more scintillators.

Effective discrimination of the waveform of this signal is important for accurate measurement of incident radiation. Therefore, we fabricated a detector (design illustrated in Figure 1, left) that combines three scintillators to discriminate between α-, β-, and γ-ray signals. This detector type is typically built up in three layers in a sandwich construction. Each layer scintillates (produces photons) as a response to particles in a particular energy range (i.e., alpha, beta, and gamma-rays). These photons are then detected by either a photomultiplier (PM), which can be either a photomultiplier tube (PMT) type or a silicon photo-multiplier (SiPM) type. These PMs then output a charge over time waveform corresponding to the incident rate of photons. Since the PM receives photons from all layers of the sandwich, there is no inherent way of separating photons, nor the subsequent charge generated from each layer. However, due to different gain, and decay properties of the phosphorescence, and fluorescence of each sandwich layer, the photons generated from each layer produce a distinct and identifiable waveform of charge over time in the PM. KNN (k-nearest neighbor) [1][2][3] is well a known method utilized for discrimination, however there are some disadvantages. In this paper, we first briefly describe our results using the KNN method, and then explain how the newly developed dSINC method mitigates some of the problems encountered with KNN.

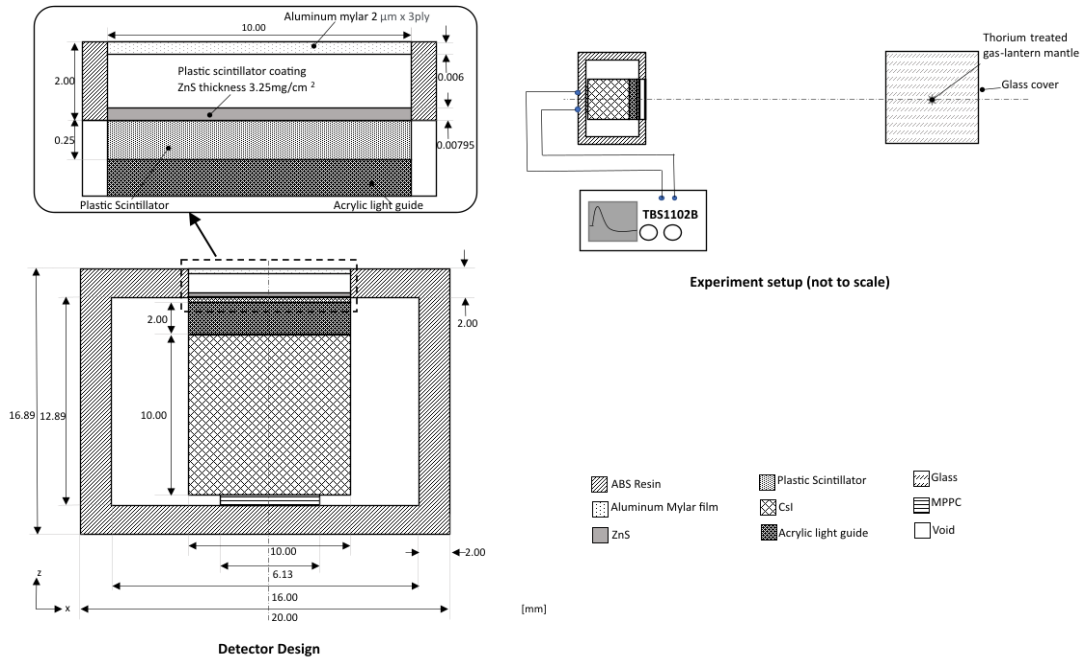

**Figure 1.** Detector design and experiment setup

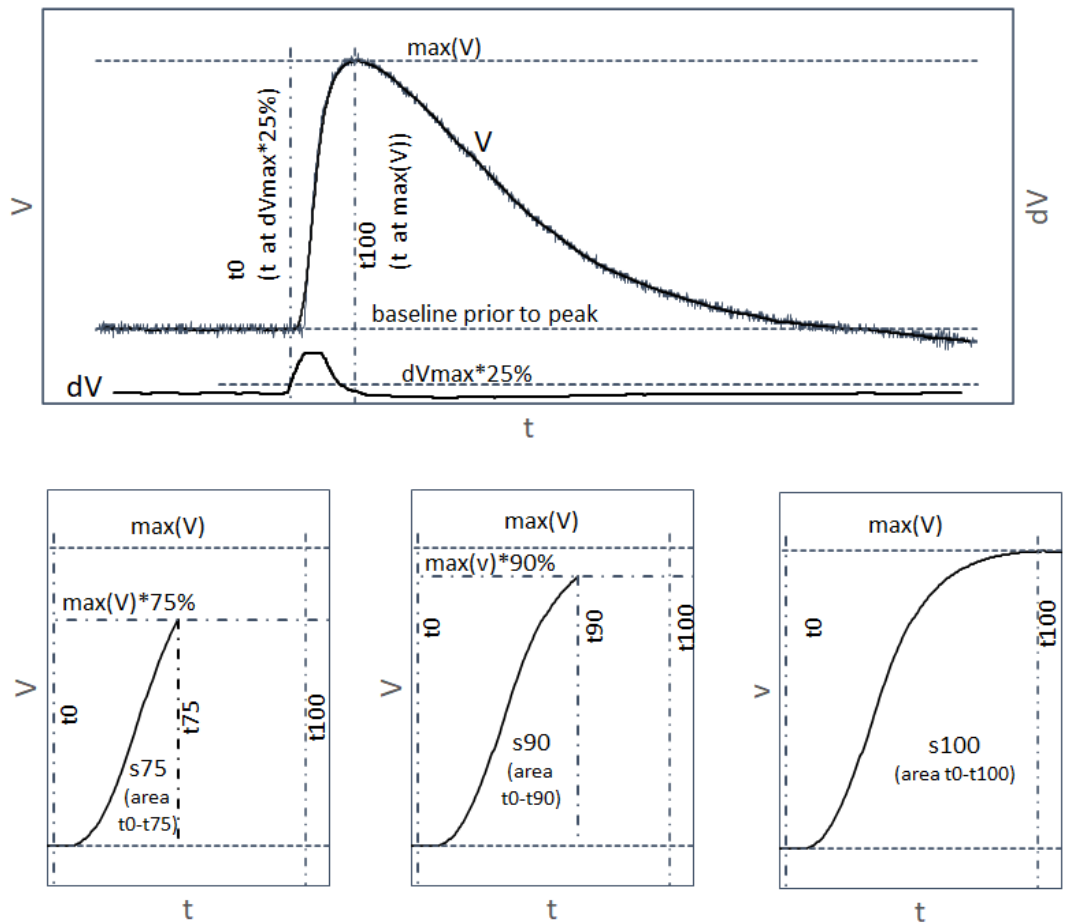

**Figure 2** Feature extraction process of KNN method

## 2 KNN, an existing technology

In our initial attempt, we employed the commonly used method of feature-extraction with subsequent KNN classification.

### 2.1 Data-sets

Both the training data, and the measurement data were gathered by connecting the detector to a Tektronix TBS1102B oscilloscope.

For the ***training data***, a set containing 372 waveforms pre-classified as alpha, 477 waveforms pre-classified as beta, and 475 waveforms pre-classified as gamma, sampled at a 250MHz sampling rate.

For the ***measurement data***, a set of 476 unclassified waveforms were gathered using the same setup, with a source consisting of a thorium treated gas lantern mantle under a glass-cover (see illustration in Figure 1, right side), sampled at a 100MHz sampling rate. For ease of comparison, this set was then up sampled to 250MHz.

### 2.2 Feature-extraction

Feature extraction is the process of measuring identifying properties of the waveform, that are uniquely identifying for each layer. Typical properties can be the max peak value of the waveform, the integrated sum of part of the waveform, or the time taken for the waveform to rise from some low fraction to a higher fraction, or even the Tail to Total (TTT) measurement. For our data, we measured the properties s75, s90, s100, t75, t90, t100, and max, see Figure 1 for an example of the process of extracting these features.

### 2.3 Training-phase

In the training-phase, the pre-classified particle-type, and the values of all the selected features were recorded for each sample in the training-data together with the corresponding classification.

### 2.4 Identification -phase

In the identification-phase KNN was used to look at these measurements in an n-dimensional Euclidean "parameter-space," and determine the k-nearest neighbors of the unknown waveforms in the measurement data. Each unknown waveform was classified according to the class of the majority of the k-nearest neighbors.

### 2.5 Parameter-search

An exhaustive parameter search was performed for each permutation of the parameter set s75, s90, s100, t75, t90, t100 (including subsets), combined with k-values ranging from 2 to 8, to find the combination of features, and k value that produced the highest quality classification.

The quality of the classification was determined by measuring the deviation of each classified sample against an ideal curve for each class. The concept of ideal curves is described further in 3.2.4, as the concept of ideal curves was the central theme for the development of the dSINC algorithm.

Table 1. Comparison of lantern test results using KNN with different features

| class | KNN s100, t100, k4 | | KNN s90, t90, k4 | |
|---|---|---|---|---|
| | count | deviation | count | deviation |
| a | 30 | 0.0937 | 77 | 0.1426 |
| b | 114 | 0.0472 | 66 | 0.0337 |
| g | 331 | 0.0591 | 332 | 0.0532 |
| sum | 475 | 0.2 | 475 | 0.2295 |
| rank | 1 | | 2 | |

**a:** alpha, **b:** beta, **g:** gamma

### 2.6 Results

In Table **1** we present a comparison of KNN method for the two feature combinations with the best performance, 1) t100, and s100 with k=4, and 2) t90, and s90 with k=4, where the former was shown to provide the best results.

### 2.7 Problems

However, there are two main problems in waveform discrimination with feature extraction and KNN, which stem mainly from two sources: noise, and peaks-piling (overlapping waveforms).

The first problem, the inherent noisiness to the system, e.g., from interferences or cross-talk affects the feature extraction step. I.e., high frequency(hf) noise may push the start and end points of the extracted features – such that the measurement values start to more closely resemble the features of other classes waveforms.

The second, Peaks-pilling, is inherent to the sandwich design, in that if several particles hit the detector in a short time span, the output will be a combined (summarized) waveform of each layer's individual waveform.

Another problem, is knowing which and how many features to include in the KNN classification. We performed an exhaustive parameter search, and found the best feature combination for the datasets used in this paper (s100, t100, k=4). Performing such a parameter search, however, requires a method to obtain an objective "figure of merit" for the fit of the entire system, this is where idea of ideal curves came from.

Furthermore, Training data is subject to potential error, not only from the above-mentioned feature extraction problems, but also from the fact that stray particles may hit the detector during the gathering of training data. To mitigate this, the training data can be manually inspected to discard samples that does not match the expected output. Besides being time consuming this potentially also induces a form of "human bias" in the training process.

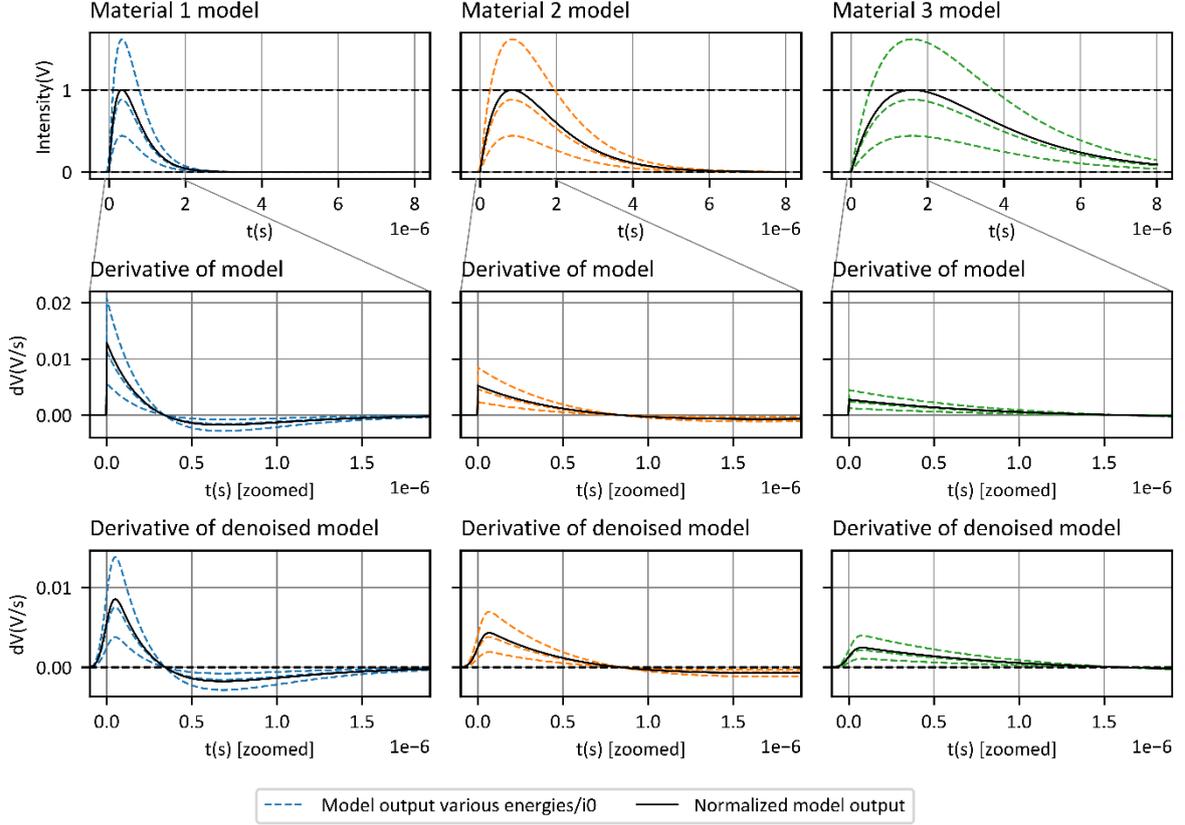

**Figure 3**. Normalization and derivative of model materials

## 3 dSINC, an alternative to KNN

### 3.1 Theoretical model

The time response of a scintillator can be modelled by the following equation [4].

$$I = I_0\left(e^{-t/\tau} - e^{-t/\tau_1}\right) \quad (1)$$

Where $t$ is time, $\tau$ is the time constant describing decay, and $\tau_1$ is the time constant describing the gain (photon population), $I_0$ the probability of photon generation.

While $\tau$, and $\tau_1$ are purely properties of the scintillator material, the probability of photon generation, $I_0$, is related both to the scintillator material and to the particle incidence energy. The time distribution of the waveform is thus a pure function of scintillator material while the amplitude of the waveform, although statistically indicative of scintillator material, also introduces overlapping and confounding properties to the measurements. The model output for 3 different artificial materials is shown in Figure 3(top row).

### 3.2 dSINC development

We developed the dSINC method to remedy the problems encountered when using KNN. In dSINC we try to learn an approximation of the normalized theoretical model by analyzing the training data. We call these learned representations ideal curves.

#### 3.2.1 Removing confounding component

First, we normalize both the training-data, and measurement-data to remove amplitude as a confounding variable component of the waveform, producing a unique identifying waveform for each material as can be seen in Figure 3 (top row, solid black line).

#### 3.2.2 Peaks piling problem

Next, we try to address the peaks piling problem by using derivative fit of the normalized data.

In Figure 3 we can see the derivative of the model output (middle row) and the derivative of the denoised model output (bottom row). In Figure 4, we can see how three concurrent waveforms combine with noise to form a combined output with piled peaks (1) and the normalized combined form (2).

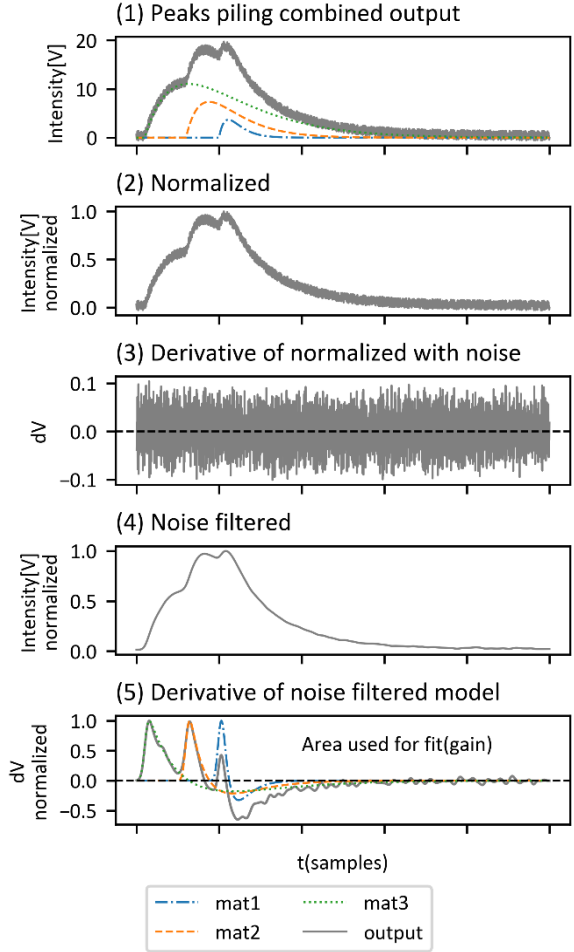

**Figure 4.** Derivative of model with overlapping(piled) peaks

### 3.2.3 Derivative of noise problem and then filtering

However, performing a derivative fit on waveforms with hf noise presents a new problem, as with a small dt, the hf noise becomes the dominant component of the derivative. The problem that we face is that the derivative of noise is just more noise (see (3) in Figure 4). This means the model data must be filtered to remove the noise, (see (4) in Figure 4).

The liquid mean filter used here is a form of iterative rolling mean, which allowed us to use a smaller window-size, better preserving the shape of the original waveforms.

$$I_{lm(j=0,i)} = I_{(t)} \quad (2)$$

$$I_{lm([0<j<l],i)} = \frac{1}{k} \sum_{i=n-2k+1}^{n+k} I_{lm(j-1,i)}$$

Where $l$ is the number of iterations, $m$ the number of samples, $I_{lm[j]}$ is the output of each iteration $j$, $n$ is the number of samples, $k$ the window size, and $i$ the current sample.

### 3.2.4 ideal curves

Ideal curves are found by taking the mean of the most representative normalized waveforms in each training data class. Where "most representative" means the waveforms with the smallest residual against the median of each class. We must also apply the same filtering algorithm to our ideal curves. This is done, not only to remove noise, but also to ensure that whatever deformation side-effect imparted to the model data in the previous step is reflected equally in the ideal curves. See Figure 4 (row 2, and 4), with the full scale, the noise filtered model is indiscernible from the original model. However, in Figure 3 (middle and bottom row) we can see the effect of noise-filtering on the model derivative curves.

### 3.2.5 Derivative of ideal curves

After noise filtering and taking the derivative, we can now see in Figure 4(5) that the combined waveform with peaks-piling contains three peaks, and that the first seems to be a good match for material 3, the second for material 2, and the last although not so clear matches up with material 1.

### 3.2.6 Synchronization & test fit

In Figure 4(5) we saw that the derivative of the individual waveforms roughly matches the derivative of the component waveforms. In the illustration the peaks line up in time due to the fact that the derivative curves are based on the time separated material peaks in Figure 4(1). However, in reality the time separation is unknown, and we need to synchronize the material ideal curves by lining up the inflection points of all material curves with those of the measurement data. (See illustration of the synchronization process in Figure 5)

## 4 Analysis for the measurement data

In this section we analyze the dSINC method trained on the csi training data-set to analyze the measurement data set LanternTest.

### 4.1 Learning ideal fit from Training Data

This section deals with learning the approximate model representation, corresponding to the scintillator material response curves from the training data. The number in parentheses corresponds to those in Figure 6

(1) Reading raw training data
Here, we are reading the training data for each group into individual data frames. Each group consisting of hundreds of data series. Each series containing 2500 samples spanning an interval of 10μs.

(2) Normalizing training data (baseline:max=0:1)
In the previously described KNN solution, we scaled each dataset according to scaling meta data parameters that were stored by the oscilloscope together with the measurements. In dSINC, no per group scaling is necessary, as each series of each group will be normalized to the range of **baseline:max = 0:1**. Baseline for each series is considered the median of the first 50 values. Values below baseline are left as negative values.

(3) Noise filtering
In the next step (4), the algorithm will synchronize the series of each group around a common point of max derivative. This means that it is important to reduce the noise of the data, or a temporary spike caused by the noise might be the point of max derivative. The simplest and fastest way to get rid of the noise is a simple rolling means function. Here we are using a window size of 50 samples. Care must be taken when selecting the window size. Too large a window will distort the shape of the curves (waveforms), too small a window size will not remove enough noise.

(4) Synchronizing
Here we find the inflection-point, or t of derivative max (tdmax), in each series and synchronize the entire group around a central tdmax. It is now becoming increasingly clear that the beta, and gamma groups respectively have some series that are misidentified between them. However, the algorithm will still work as long as there aren't too many misidentified series in a group.

(5) Finding ideal fit(curve)
An eager learning algorithm could at this point attempt to fit the model from Equation 1, to determine $\tau$, and $\tau_1$ for each group of training data. Here we take the more general approach of using ideal fitting curves, that is not tied to a specific model. Technically the ideal fitting curve should be the curve which minimizes the residual square sum error against a group of training data. This curve can be found by any iterative type solver. But here we are foregoing the solver and going with a simpler approach which seems to work well enough. The ideal curve for each group, icg is found by Algorithm A-2. [Note] The original training data includes incorrect classifications (most obvious in the beta-class). In step 5 only the 10% most representative samples are selected to represent the group as basis for the ideal curve.

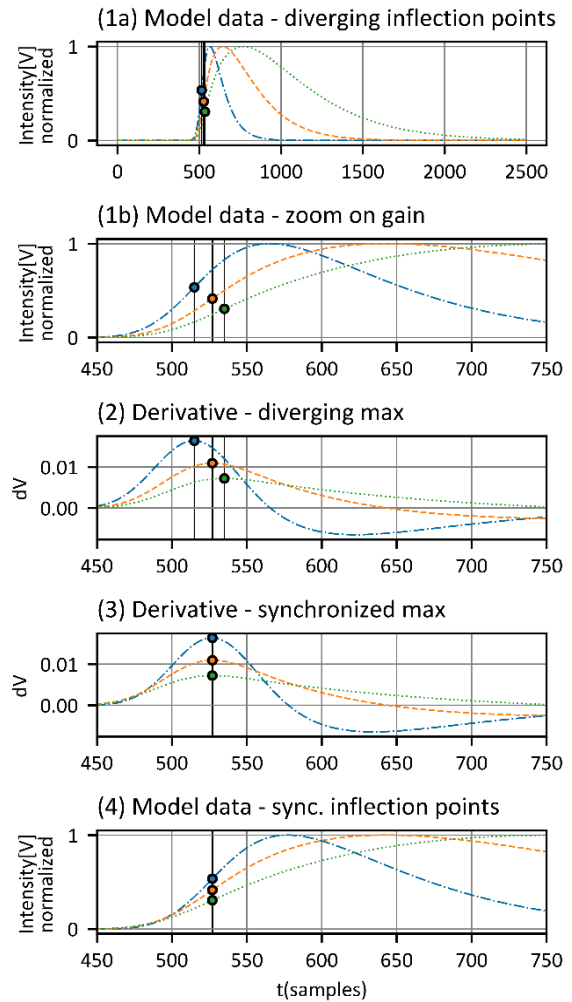

**Figure 5.** Synchronization of curves around median inflection point

### 4.2 Examining Measurement Data

This section explains how to examine 'measurement data'. Number in parentheses corresponds to those in Figure 7.

(1) Read Measurement Data
The first step of the fitting process is to read the measurement data to be fitted. In this case, we are skipping two broken datasets, which if included would distort the visualizations.

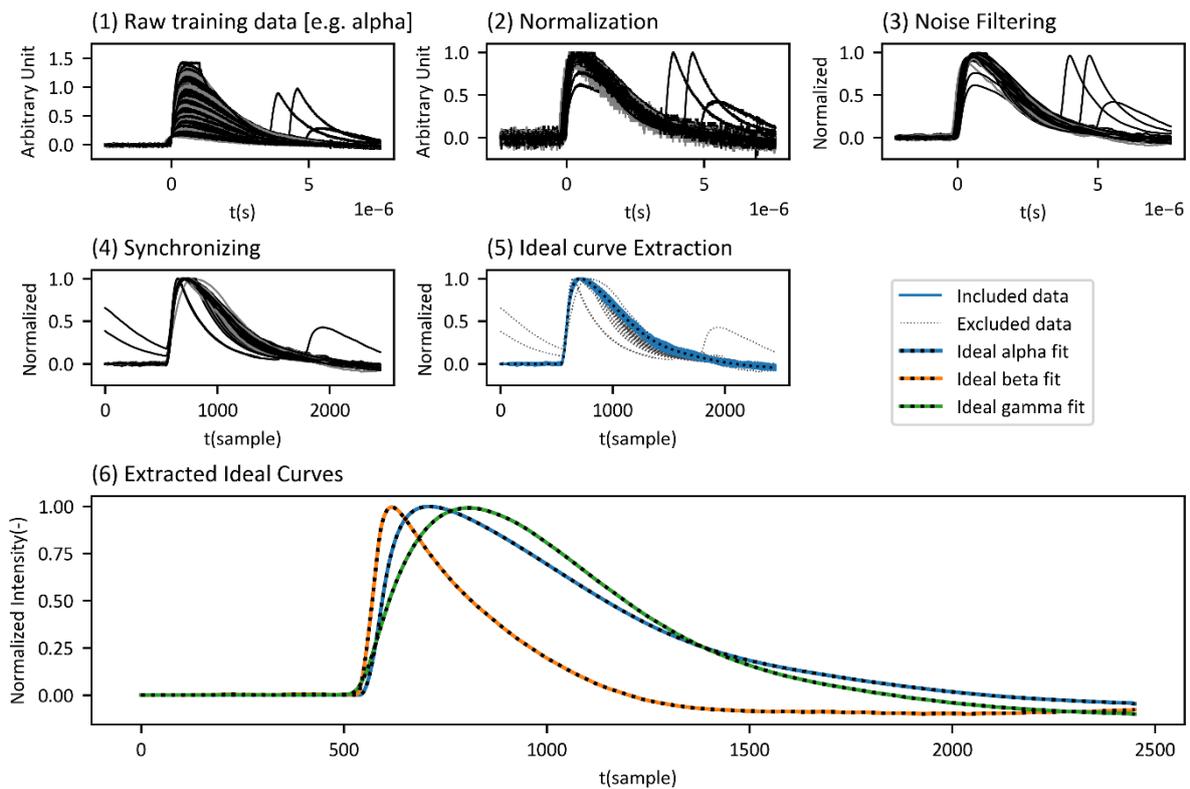

**Figure 6.** Images of ideal curve extraction process of training data

(2) Resampling Measurement Data
In this particular case, the sampling rates of training data, and measurement data does not match, so we resample each series with a factor of 2.5 using linear interpolation. This step is not needed if sampling rates are already matching.

(3) Normalizing Measurement Data
In this step we normalize the measurement data to ***baseline:max = 0:1***.

(4) Removing noise from Measurement Data
In this step we remove noise and renormalize the data to ***baseline:max = 0:1*** again.

(5) Synchronizing Measurement Data
Just like with the training data we can synchronize measurement data at tdmax. It is now possible to start looking at a comparison between the measurement data and the ideal curves.

(6) Zooming in on area of comparison
In the case of the KNN algorithm, it only compares waveforms on properties that are calculated on the 'gain' section of the waveform. dSINC also chooses to only fit on the 'gain' part of the waveform for two reasons; 1) The gain section is shorter than the tail(decay). This means there is less probability of interference from a second overlapping peak. 2) It makes it easier to make a fair comparison between KNN and dSINC. In the graph in Figure 7 (6), and (7) we zoom in on the 'gain portion of the ideal curves. We can now see that most curves seem to fit either beta, or gamma, with few or no series fitting to alpha.

(7) Renormalizing area of comparison and classification
In the previous step we could see a problem. Many of the series that look like potential fit for gamma seem offset in Y from the gamma fit line (with square symbols). This is most likely because the original normalizes to baseline operations were performed against a baseline taken from the beginning of the series (which could potentially have been noisy). Now that we have identified the area of comparison, we can however repeat this normalization process on the selected area only. Finally, see the results in Figure 7

## 4.3 Comparison between dSINC and KNN

**Table 2** compares the results of dSINC with the results from KNN using the best ranking feature set. It evidently shows the advantage of the newly developed dSINC method compared to conventional KNN. dSINC can also classify data as 'uncertain' when the confidence score of a sample is low. dSINC scores are presented both with and without the option to allow the class (**~**) for measurement data with a low confidence. In either case, the error sum of deviation from the ideal curve is lower for dSINC.

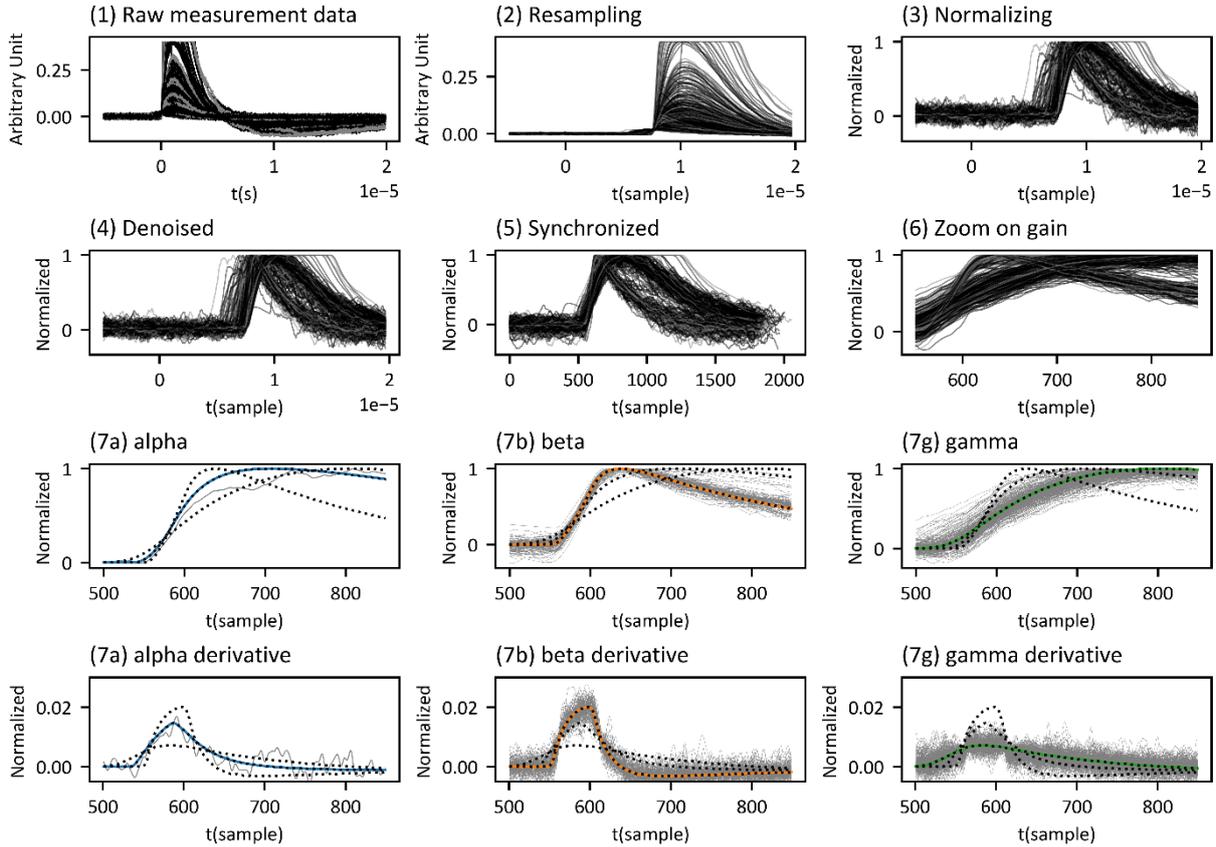

**Figure 7.** Process to examine 'measurement data'

Table 2. Comparison of lantern test results between dSINC and best KNN

| class | dSINC allow unclassified (~) | | dSINC no unclassified (~) | | KNN s100, t100, k4 | |
|---|---|---|---|---|---|---|
| | count | deviation | count | deviation | count | deviation |
| a | 1 | 0.0426 | 3 | 0.0799 | 30 | 0.0937 |
| b | 135 | 0.0396 | 138 | 0.0420 | 114 | 0.0472 |
| g | 326 | 0.0498 | 334 | 0. 0511 | 331 | 0.0591 |
| ~ | 13 | - | - | - | - | - |
| sum | 475 | 0.132 | 475 | 0.1730 | 475 | 0.2 |
| rank | 1 | | 2 | | 3 | |

**a:** alpha, **b:** beta, **g:** gamma, **~:** uncertain

## 5 Conclusions

In the KNN cases compared in Table **1**, there is a big discrepancy in the resulting alpha and beta counts even between the two best cases based on feature set s100, t100 (alpha count 30, beta count 114) and the case based on feature set s90, t90 (alpha count 77, beta count 66), but the KNN method itself provides no indication as to which feature set is the most accurate.

Using a figure of merit based on sample residual (deviation in the Table **1**, and Table **2**), against a training data class-median (corresponding to ideal curve in dSINC) indicates that s100, t100 was the better option. But the figure of merit cannot be calculated until KNN for each feature-set already has been completed.

dSINC however have the ideal curve, and its residual as an integral part of the algorithm. And the synchronized and normalized measurement data can easily be compared against the ideal curves for verification (see Figure 7). For waveform-discrimination of the presented test-case, dSINC classification outperforms KNN classification. The deviation score of dSINC case with "no unclassified" shows a better deviation score (0.173) compared to the best KNN case (0.2) ( Table **2**, column 2 vs. column 3). The dSINC deviation score is further improved to 0.132 (Table **2**, column 1), if we allow dSINC to exclude 'uncertain' (~) measurement data from the deviation score.

dSINC achieves this result without the need of feature extraction or configuration.

## Acknowledgements

This work was partially carried out in a subsidy program of "Project of Decommissioning and Contaminated Water Management", entitled "Development of Technologies for Work Environmental Improvement in Reactor Building (Development of Exposure Reduction Technologies by Digitalization of Environment and Radioactive Source Distribution)".

# Appendix

## Training phase

The training data phase is described in high level overview in (Algorithm A-1). First the training data of each class is read groupwise into list TD['a','b','g'] for alpha, beta, and gamma respectively. Each TD group entry is a data frame where each column represents a training data sample series. Next, for each group **g** of the training data, the ideal representation **icg** is found by the function **ideal_repr** (Algorithm A-2), noise filtering is applied to the ideal representation **icg**, and stored as ideal curve for the group in **IC[g]**.

**Algorithm A-1** Training Phase - high level overview

| |
|---|
| Read and store training data groupwise into **TD**[ **'a','b','g'** ] |
| for each group **g** in **TD**: |
|    **icg** ← ideal_repr( **TD[g]** )    //Algorithm A-2 |
|    **IC[g]** ← liquid_mean( **icg**, n=50, win=3 ) //Equation 2 |

To find the ideal representations, or ideal curves, of each group of training data, the function **ideal_repr** (Algorithm A-2) is used. Here we calculate the $L^2$-norm of each series and store to list **N**, next we take the median **m** of all norms in **N**. Having a median norm, we can calculate a residual **R** for each norm in **N** describing the distance **N-m**.
In the next step we find a subset of size **k** of **g** by taking the **k** number of entries with the smallest **r** and storing them to **s**. We now calculate the ideal curve **ic** by taking the mean of each series **s** at each sample point **t**.

**Algorithm A-2** ideal_repr – find ideal waveforms

| |
|---|
| Let **g** be a data frame of preclassified training data, where each column represent a waveform data series in the training dataset. |
| **n** ← len( **g** )    // number of entries in group |
| **N**[**1..n**] ← l2norm( **g**[**1..n**] )   // the $L^2$-norm of each entry |
| **m** ← median( **N** )    // median of all norms |
| **R** ← abs( **N - m** )   // residual (distance to median) |
| **s** ← ksmallest( **g** ,**r,** k=**n***10%)  // subset s, k best residuals |
| **ic** ← mean( **s[t]** )    // ideal curve from mean of s |

## Classification phase

The classification phase is described in high level overview in (Algorithm A-3). First each data series in the measurement data is read into **md**, and the liquid mean noise filtering (Equation 2) is applied to each series. Next non sequential dSINC (Algorithm A-4) is used to find the fit of each wave form data series in **md** against each ideal curve in **IC** and the best fit is stored in **dSF**. And finally, the distance **dist**, **score**, and **class** of each best fit **dSF** is found by **dsinc_classify** (Algorithm A-5).

**Algorithm A-3** Classification Phase – high level

| |
|---|
| **md**[**1..n**] ← read_measurement_data( ) |
| **md**[**1..n**] ← liquid_mean(**md**[**1..n**], **n**=50, **win**=3)   // Algorithm 3-5 |
| dSF ← nsq_dsinc( md[1..n], IC )    // Algorithm 3-4 |
| [**dist, score, class** ] ← dsinc_classify( **dSF**, **limit**=-0.0003) |

The non-sequential dsinc fit algorithm fits measurement data where the waveforms are captured independently in separate series. nsq_dsinc first normalizes the values in the measurement data X, and ideal curves IC and store to XN, and INC. The values are normalized to the range 0:1, with the base line 0 set to the median of the first 50 samples in the data (any negative values are kept).
Next the derivative (dt = sample rate) is calculated for XN, and INC and stored in dXN, and dINC. In lines 5-12, dINC, INC, and dXN are synchronized around their point of highest derivative value and stored in dSINC, SINC, and dSXN.
In lines 13-14, the area of comparison where SINC > limit(=0.001) found. And finally in lines 15-20, the residual of the area of comparison of dSXN-dSINC is calculated for each measurement data series against the ideal curve of each group g in GC.

## Algorithm A-4 nsq_dsinc - non-sequential dSINC fitting

Let **IC** be the filtered ideal curves
Let **X** be the clean measurement data
Let **nc** be the number of groups(classes) in **IC**
Let **limit** be a low number between 0 and 1, default **0.001**

// Normalize ideal-curves, and measurement data to baseline
**XN** ← normalize( **X**, baseline=median( **X**[0:50] ) )
**INC** ← normalize( **IC**, baseline=median( **IC**[0:50] ) )
// Get derivative of ideal-curves, and measurement data
**dXN** ← derivative( **XN**, **dt**=min_sample )
**dINC** ← derivative( **INC**, **dt**=min_sample )
// Find alignment centers by median of windowed means
dinc_alignments ← idxmax( wmean( dINC ))
**dxn_alignments** ← idxmax( wmean( **dXN** ))
dinc_align_center ← median( dinc_alignments )
dxn_align_center ← median( dxn_alignments )
// Synchronize sample data
dSINC ← shift( dINC, dinc_align_center - dinc_alignments )
SINC ← shift( INC, dinc_align_center - dinc_alignments )
dSXN ← shift( dXN, dxn_align_center - dxn_alignments )
// Find time range samples to compare
start ← idx( SINC > limit )[0]
end ← idx_max( SINC )
// Calculate residuals
dSF = frame(zeros(end-start+1, nc)
for each group **g** in **IC**:
   ic ← dSINC[ g ][ start[ g ]:end[ g ] ]
   ds ← dSXN[ start[ g ]:end[ g ] ]
   r ← ( sum( abs( ds-ic )) / ( end-start ))
   dSF[g] ← r

## Algorithm A-5 dsinc_classify - dSINC classification

Let **dSF** be the fitting score for the measurement data
Let **limit** be the minimum rss (negative) score considered classifiable
Let **G** be the list of groups(classes) in **dSINCfit**

// Get best dSINC distance(residual)
**dist** ← min( **dSF**, axis=1 ) / max(max( **dSF**, axis=1 ) )
**idx_best_rss** ← idxmin( **dist**, axis=1 )
// Calculate score relative to central distance
**central_rss** ← median( **dSF**, axis=1)
score ← min( dSF – central_rss, axis=1)
// Classify according to best score, where score < limit
**class** ← ['~']*len(**idx_best_rss**)     // ~ = unknown
class[ score < limit ] ← G[ idx_best_rss ][ score < limit ]